\documentclass[aps,prd,twocolumn,superscriptaddress,showpacs,showkeys]{revtex4-1}
\usepackage{amsmath,amssymb,graphicx,bm,braket,nicefrac}

\newcommand{\darmstadt}{\affiliation{Institut f\"ur Kernphysik,
Technische Universit\"at Darmstadt, D-64289 Darmstadt, Germany}}
\newcommand{\emmi}{\affiliation{ExtreMe Matter Institute EMMI, 
GSI Helmholtzzentrum f\"ur Schwerionenforschung GmbH, 
D-64291 Darmstadt, Germany}}
\newcommand{\purdue}{\affiliation{Department of Physics,
Purdue University, West Lafayette, IN 47907, USA}}
\newcommand{\zurich}{\affiliation{Physics Institute, 
University of Z\"{u}rich, Winterthurerstr.~190, CH-8057, Switzerland}}

\usepackage{color}

\begin{document}

\title{Signatures of Dark Matter Scattering Inelastically Off Nuclei}

\author{L.\ Baudis}
\email{laura.baudis@physik.uzh.ch}\zurich
\author{G.\ Kessler}
\email{gaudenz.kessler@physik.uzh.ch}\zurich
\author{P.\ Klos}
\email{pklos@theorie.ikp.physik.tu-darmstadt.de}\darmstadt\emmi
\author{R.\ F.\ Lang}
\email{rafael@purdue.edu}\purdue
\author{J.\ Men\'endez}
\email{javier.menendez@physik.tu-darmstadt.de}\darmstadt\emmi
\author{S.\ Reichard}
\email{sreichar@purdue.edu}\purdue
\author{A.\ Schwenk}
\email{schwenk@physik.tu-darmstadt.de}\emmi\darmstadt

\begin{abstract}
Direct dark matter detection focuses on elastic scattering of dark
matter particles off nuclei. In this study, we explore inelastic
scattering where the nucleus is excited to a low-lying state of
$10-100$\,keV, with subsequent prompt de-excitation. We calculate the
inelastic structure factors for the odd-mass xenon isotopes based on
state-of-the-art large-scale shell-model calculations with chiral
effective field theory WIMP-nucleon currents. For these cases, we find
that the inelastic channel is comparable to or can dominate the
elastic channel for momentum transfers around $150$\,MeV. We calculate
the inelastic recoil spectra in the standard halo model, compare these
to the elastic case, and discuss the expected signatures in a xenon
detector, along with implications for existing and future
experiments. The combined information from elastic and inelastic
scattering will allow for the determination of the dominant interaction channel
within one experiment. In addition, the two channels probe different
regions of the dark matter velocity distribution and can provide
insight into the dark halo structure. The allowed recoil energy domain
and the recoil energy at which the integrated inelastic rates start to
dominate the elastic channel depend on the mass of the dark matter
particle, thus providing a potential handle to constrain its mass.
\end{abstract}


\maketitle

\section{Introduction}

Astrophysical evidence indicates that the milky way disc is embedded
into a non-baryonic dark matter halo~\cite{Nesti:2013uwa}, which could
be made of weakly interacting massive particles (WIMPs)~\cite{Feng:2010gw}.
The WIMP dark matter hypothesis is testable, as WIMPs may be detected
directly by scattering off nuclei in low-background underground
detectors, indirectly by observing annihilation products above
astrophysical backgrounds, and by producing them at the
LHC~\cite{Bertone:2010zz}. In direct detection experiments, only the
elastic scattering channel, with an exponential nuclear recoil energy
spectrum, is usually exploited~\cite{Baudis:2012ig}.

Another avenue to direct detection is to observe inelastic
WIMP-nucleus scattering, inducing transitions to low-lying excited
states~\cite{Ellis}. The experimental signature is a nuclear recoil
together with the prompt de-excitation photon. In odd-mass nuclei with
low-lying excited states, a galactic WIMP with sufficient kinetic
energy can induce inelastic excitations. Xenon is an excellent target
material because natural xenon contains the odd $^{129}$Xe and
$^{131}$Xe isotopes with abundances of 26.4\% and 21.2\%,
respectively. The excitation energies of the lowest-lying states are
reasonably low, with a $\nicefrac{3}{2}^+$ state at 39.6\,keV above
the $\nicefrac{1}{2}^+$ ground state in $^{129}$Xe and a
$\nicefrac{1}{2}^+$ state at 80.2\,keV above the $\nicefrac{3}{2}^+$
ground state in $^{131}$Xe. The nuclear decays are electromagnetic M1
and E2 with half-lives of 0.97\,ns and 0.48\,ns,
respectively. Searches for inelastic scattering have been performed in
the past~\cite{Ejiri,Belli,Avignone,Bernabei:2000qn}.

Here, we show that inelastic WIMP-nucleus scattering in xenon is
complementary to elastic scattering for spin-dependent interactions,
as the inelastic channel dominates the integrated spectra above
$\sim$\,10\,keV energy deposition, depending on the WIMP mass. This
aspect not only provides a consistency check for a xenon experiment,
but in the case of dark matter detection via this channel, it would
offer a clear indication for the spin-dependent nature of the
fundamental interaction.

This paper is organized as follows. In
Sect.~\ref{sec:structure_factors}, we calculate the structure factors
for inelastic WIMP-nucleus scattering, which is dominated by
spin-dependent interactions for the relevant low-lying excited
states. The kinematics of inelastic scattering is described in
Sect.~\ref{sec:kinematics}. In Sect.~\ref{sec:signatures}, we explore
the expected inelastic signature in a xenon dark matter detector and
compare it with the signal region for elastic nuclear
recoils. Finally, we discuss in Sect.~\ref{sec:implications} the
implications of this signature for current and future dark matter
detectors such as LUX, LZ, XENON, XMASS, and DARWIN, and conclude.

\section{Inelastic structure factors}
\label{sec:structure_factors}

\subsection{Nuclear structure and WIMP-nucleon currents}

The calculation of inelastic WIMP scattering off nuclei requires a
reliable description of the structure of the initial and final nuclear
states as well as the WIMP-nucleon currents. Because low-lying
transitions occur in odd-mass isotopes with different spins of the
ground and excited states, spin-dependent WIMP scattering generally
dominates for the inelastic channel (see Ref.~\cite{Engel:1999kv},
which studied the spin-independent case). We therefore consider only
spin-dependent WIMP-nucleon interactions.

We perform state-of-the-art large-scale nuclear structure calculations
for $^{129}$Xe and $^{131}$Xe using the shell-model code
ANTOINE~\cite{Caurier:2004gf} and the GCN5082
interaction~\cite{CaurierPRL,MenendezNPA} in the $0g_{7/2}, 1d_{5/2},
1d_{3/2}, 2s_{1/2}$, and $0h_{11/2}$ valence space on top of a
$^{100}$Sn core as in Refs.~\cite{Menendez,Klos}. For $^{129}$Xe, the
number of particle excitations into the last three orbitals was
limited to three; for $^{131}$Xe, we perform an exact
diagonalization. These present the largest valence spaces with nuclear
interactions that have been tested in nuclear structure and decay
studies. The resulting energy spectra show a very good overall
agreement with experimental data~\cite{Menendez}, and the spins of the
ground and first excited states are correctly predicted, a clear
improvement with respect to previous work~\cite{Toivanen:2009}. The
calculated excitation energies for the first excited states in
$^{129}$Xe and $^{131}$Xe, 107\,keV and 37\,keV, are in reasonable
agreement with the experimental values, given these very small
energies. Note that for the inelastic WIMP-nucleus cross sections, we
will use the experimental excitation energies, so the calculation uses
as input only the wave functions of the nuclear states.

While WIMPs interact with quarks, at the momentum scales relevant to
nuclei and WIMP-nucleus scattering, the relevant degrees of freedom
are nucleons and pions. In this regime, chiral effective field theory
(EFT) provides a systematic expansion in powers of momentum $Q$ for
the coupling of WIMPs to nucleons based on the symmetries of QCD. At
leading orders $Q^0$ and $Q^2$, chiral EFT predicts one-body (1b)
currents given by~\cite{Menendez,Klos}
\begin{align}
\sum_{i=1}^A {\bf J}^3_{i,{\rm 1b}} &= \sum_{i=1}^A \frac{1}{2} \biggl[ 
a_0 \, {\bm \sigma}_i \nonumber \\[1mm]
&\quad + a_1 \tau_i^3 \biggl( \frac{g_A(p^2)}{g_A} \, {\bm \sigma}_i
- \frac{g_{P}(p^{2})}{2 m g_A} \,
({\bf p} \cdot {\bm \sigma}_{i}) \, {\bf p} \biggr) \biggr] \,,
\label{1bcurrents}
\end{align}
where the sum is over all $A$ nucleons in the nucleus, and we
consider spin-$\nicefrac{1}{2}$ WIMPs. These 1b currents are similar to the
phenomenological currents that have been used in spin-dependent WIMP
scattering off nuclei~\cite{Engel:1992bf}. Here, $a_0$ and $a_1$
are the isoscalar and isovector WIMP-nucleon couplings;
$\bm{\sigma}_i$ and $\tau^3_i$ are spin and isospin matrices; and
${\bf p} = {\bf p}_i-{\bf p}_f$ denotes the momentum transfer from
nucleons to WIMPs. $g_A(p^{2})$ and $g_P(p^{2})$ are the axial and
pseudoscalar couplings including $Q^2$ corrections given in
Ref.~\cite{Menendez}, $g_A(0)=g_A$ and $m$ is the nucleon mass.

In addition to the coupling through 1b currents, chiral EFT predicts
two-body (2b) currents at order $Q^3$, where WIMPs couple to two
nucleons~\cite{Menendez}. Two-body currents are quantitatively
important in medium-mass and heavy nuclei because the momentum $Q$
involved is not only sensitive to the momentum transfer but also to
the typical momenta of nucleons in nuclei, which is higher (of order
of the Fermi momentum)~\cite{2bcurrents}. The normal-ordered 1b part
of the dominant long-range 2b currents is derived summing over
occupied states in a spin-isospin symmetric reference state, which we
take as a Fermi gas~\cite{2bcurrents,Menendez,Klos}. This is expected
to be a very good approximation due to phase-space
considerations~\cite{Friman} and allows direct comparison to
Eq.~\eqref{1bcurrents}. The inclusion of the long-range 2b currents
leads to additional contributions to the isovector axial and
pseudoscalar currents. Combining the 1b and 2b currents to order $Q^3$
gives~\cite{Klos}
\begin{align}
{\bf J}^3_{i,{\rm 1b+2b}} &= \frac{1}{2} \, a_1 \tau_i^3 
\biggl[ \biggl(\frac{g_A(p^2)}{g_A}+\delta a_1(p)\biggr) {\bm \sigma}_i
\nonumber \\[1mm]
&\quad+ \biggl(-\frac{g_P(p^{2})}{2 m g_A} + \frac{\delta a_1^P(p^2)}{p^2}
\biggr) ({\bf p} \cdot {\bm \sigma}_{i}) \, {\bf p} \biggr] \,,
\label{J3}
\end{align}
with a momentum- and density-dependent renormalization of the axial-vector
and pseudoscalar parts, $\delta a_1(p)$ and $\delta a^P_1(p)$, respectively,
\begin{align}
\delta a_1(p) &= -\frac{\rho}{F^2_\pi} \, 
\biggl[ \frac{1}{3} \Bigl(c_4+\frac{1}{4m}\Bigr) 
\Bigl[3I_2^{\sigma}(\rho,p)-I_1^{\sigma}(\rho,p) \Bigr] \nonumber \\[1mm]
&\quad+\frac{1}{3}\Bigl(-c_3+\frac{1}{4m}\Bigr) 
I_1^{\sigma}(\rho,p) - \Bigl( \frac{1+\hat{c}_6}{12m} 
\Bigr) I_{c6}(\rho,p) \biggl] , \\
\delta a_1^P(p) &= \frac{\rho}{F^2_\pi} \biggl[ 
\frac{-2c_3 p^2}{m^2_\pi+p^2} \nonumber \\[2mm]
&\quad +\frac{c_3+c_4}{3} \, I^P(\rho,p) 
- \frac{1+\hat{c}_6}{12m} \, I_{c6}(\rho,p) \biggr] \,,
\end{align}
with density $\rho$, pion mass and decay constant $m_\pi$, $F_\pi$,
and chiral EFT low-energy couplings $c_3$, $c_4$, $\hat{c}_6$. The
values for the couplings and the functions $I_1^{\sigma}(\rho,p)$,
$I_2^{\sigma}(\rho,p)$, $I^P(\rho,p)$, and $I_{c6}(\rho,p)$ (due to
integrals in the exchange terms) are given and explained in detail in
Ref.~\cite{Klos}.

The uncertainties in chiral 2b currents are dominated by the
uncertainties of the $c_3$ and $c_4$ coupling. Considering a
conservative range of values accepted in the literature leads to
$\delta a_1(0)=-(0.14-0.32)$~\cite{Klos} (the dependence on $p$ is
very weak), so that 2b currents reduce the axial part of the
WIMP-nucleon currents. On the other hand, $\delta a_1^P(m_{\pi})
=0.23-0.54$~\cite{Klos}, leads to an enhancement of the
pseudoscalar part. This enhancement vanishes at $p=0$, due to
the pseudoscalar nature, but increases with momentum transfer.

\subsection{Structure factors}

Based on the initial ($i$) and ($f$) final states, we calculate the
structure factors of inelastic, spin-dependent WIMP-nucleus
scattering. The structure factor $S_A(p)$ receives contributions from
longitudinal (${\mathcal L}^5$), transverse electric (${\mathcal
T}^{\mathrm{el}5}$) and transverse magnetic (${\mathcal T}^{\mathrm{mag}5}$)
multipoles~\cite{Klos}:
\begin{align}
S_A(p) &= \sum_{L \geqslant 0} \bigl|\bra {J_f}\!|{\mathcal L}_L^5|\!
\ket{J_i}\bigr|^2 \nonumber \\
&\quad +\sum_{L \geqslant 1} \Bigl(\bigl|\bra{ J_f}\!|
{\mathcal T}_L^{\mathrm{el}5}|\!\ket{J_i}\bigr|^2
+\bigl|\bra{ J_f}\!|{\mathcal T}_L^{\mathrm{mag}5}|\!
\ket{ J_i}\bigr|^2\Bigr) \,,
\end{align}
where the detailed expressions for the multipoles in terms of the
1b+2b currents of Eq.~\eqref{J3} are given in Ref.~\cite{Klos} (see
Eqs.~(22)--(24)). Due to parity conservation, only odd/even $L$
multipoles contribute for the electric/magnetic multipoles when the
inelastic transition is between nuclear states with the same parity
(as is the case for xenon). While magnetic multipoles do not
contribute to elastic scattering because of time-reversal symmetry,
they do contribute to the inelastic case~\cite{Klos}.

\begin{figure}[!htb]
\begin{center}
\includegraphics[width=0.975\columnwidth,clip=]{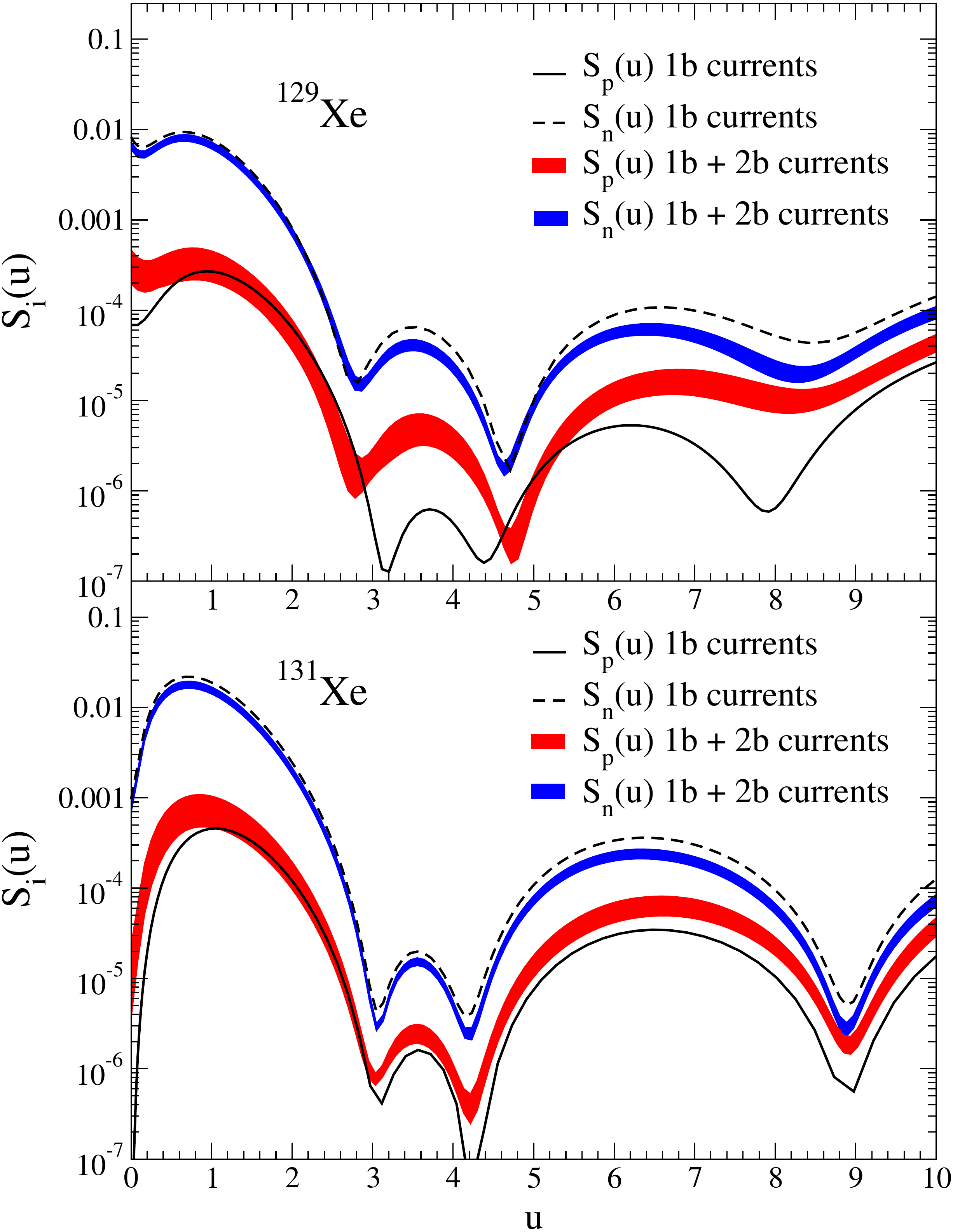}
\end{center}
\caption{(Color online) Inelastic structure factors $S_{p}$ (solid
lines) and $S_{n}$ (dashed) for $^{129}$Xe (top panel) and $^{131}$Xe
(bottom panel) as a function of $u=p^2b^2/2$. The harmonic-oscillator
length is $b= 2.2853$\,fm for $^{129}$Xe and $b= 2.2905$\,fm for $^{131}$Xe.
Results are shown at the one-body~(1b) current
level and including two-body~(2b) currents. The estimated theoretical
uncertainty in the 2b currents is given by the red ($S_{p}(u)$) and
blue ($S_{n}(u)$) bands.\label{fig:structurefunctions}}
\end{figure}

The inelastic structure factors for WIMP scattering off $^{129}$Xe and
$^{131}$Xe are shown in Figure~\ref{fig:structurefunctions}. Our
results are presented in terms of the structure factors with
``proton-only'' ($a_0=a_1=1$) and ``neutron-only'' ($a_0=-a_1=1$)
couplings, $S_p(u)$ and $S_n(u)$, as a function of the dimensionless
momentum transfer $u = p^2b^2/2$ with harmonic-oscillator length
$b$. These coupling choices are more sensitive to protons and
neutrons, respectively.  Because xenon has an even number of protons,
practically all the spin, and therefore the nuclear spin response, is
dominated by neutrons, which also dominate the ``proton-only''
structure factors at low momentum transfer through the proton-neutron
strong interaction in 2b currents. For an extended discussion see
Ref.~\cite{Klos}. 
Data files of both the elastic~\cite{Menendez, Klos} and inelastic structure functions are available as Supplemental Material in the Appendix.

\begin{figure}[!htb]
\begin{center}
\includegraphics[width=0.975\columnwidth,clip=]{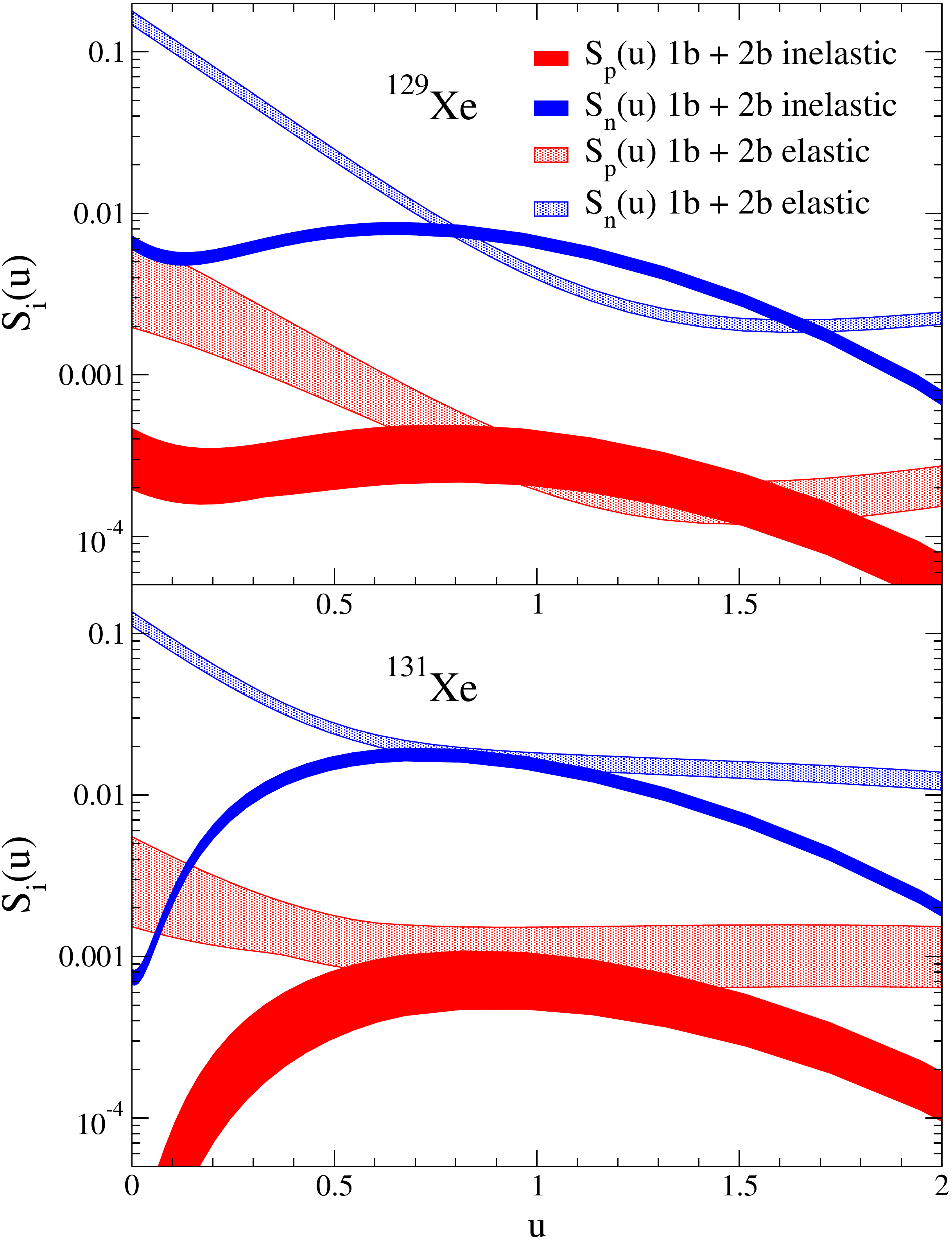}
\end{center}
\caption{(Color online) Comparison of the inelastic structure factors from Figure~\ref{fig:structurefunctions} to elastic counterparts from Ref.~\cite{Menendez, Klos}. Data files with these structure factors are available as supplementary material. \label{fig:comparisonsf}}
\end{figure}

In Figure~\ref{fig:comparisonsf} we compare the inelastic structure factors to the elastic ones from Refs.~\cite{Menendez, Klos} for the experimentally relevant region. At $p=0$, both inelastic
$S_p(u)$ and $S_n(u)$ are significantly smaller than their elastic
counterparts, a factor~$10$ for
$^{129}$Xe and more than two orders of magnitude for
$^{131}$Xe. However, for both isotopes, we observe a maximum in the
inelastic structure factors at low momentum transfer, which was also
found in previous calculations~\cite{Toivanen:2008,Toivanen:2009}.
This feature is absent in elastic scattering, where the maximum always
occurs at $p=0$ and then sharply decreases~\cite{Menendez,Klos}. As a
result, for $u=1-2$ ($p=125-175$\,MeV), the inelastic channel is
comparable to the elastic one. This is relevant because it is within
the range of allowed momentum transfers in inelastic scattering (see
Sect.~\ref{sec:kinematics}). In particular, for $^{129}$Xe, the
inelastic structure factors even dominate their elastic counterparts.
For $^{131}$Xe, the inelastic structure factors are slightly smaller
but comparable to the elastic case.

Similar results comparing inelastic and elastic scattering were found
in Ref.~\cite{Toivanen:2009}, although their structure factors are
very different to ours. For example, in $^{129}$Xe, the
inelastic structure factor is always decreasing in
Ref.~\cite{Toivanen:2009}; for $^{131}$Xe, the suppression of the
inelastic response at $p=0$ is an order of magnitude smaller than in
our case, which is compensated by a rather flat momentum-transfer
dependence; and both elastic and inelastic structure factors for
$^{131}$Xe in Ref.~\cite{Toivanen:2009} are larger than ours at $u
\sim 1$. These differences are due to the well-tested interactions and
the smaller truncations in the present calculations (see also
Ref.~\cite{Menendez}).

To understand the behavior at $p=0$ and the maxima in the inelastic
structure factors compared to the elastic case, we have to consider
the spin and parity of the ground state of $^{129}$Xe,
$\nicefrac{1}{2}^+$, and its first excited state,
$\nicefrac{3}{2}^+$. The situation is reversed in $^{131}$Xe with a
$\nicefrac{3}{2}^+$ ground state and a $\nicefrac{1}{2}^+$ first
excited state. Therefore, only $L=1,2$ multipoles contribute to
inelastic scattering. For elastic scattering, where the magnetic
multipoles do not contribute, the allowed multipoles are $L=1$ for
$^{129}$Xe and $L=1,3$ for $^{131}$Xe. At $p=0$, only $L=1$
multipoles contribute to the scattering, and only orbitals of the
initial and final states with the same orbital angular momentum $l$
can be connected~\cite{Klos}. Because these contributions are
naturally maximal in elastic scattering, but smaller when nucleons are
excited into different $l$ orbitals, there is a strong reduction in
the inelastic structure factors at $p=0$ compared to the elastic case.

At $p>0$, all multipoles contribute, and inelastic scattering is no
longer suppressed, leading to a maximum in the structure factor.
Moreover, because magnetic multipoles with $L=2$ only contribute to
inelastic scattering, this response will be enhanced with respect to
the elastic case. The extent of the suppression at $p=0$ and the
maximum of the inelastic structure factor depend on the nuclear
structure details of the states involved and on the relative
contribution of the magnetic multipoles. For $p=0$, considering 1b
currents only, the inelastic structure factors can be related to the
spin part of magnetic dipole $B(M1)$ transitions. However, a precise
fine-tuning to experimental $B(M1)$ values probes different physics
because, first, 2b currents are different for electromagnetic and
WIMP-nucleon interactions and, second, magnetic multipoles vanish at
$p=0$, while they are crucial for the inelastic structure factors.

\section{Kinematics of inelastic scattering}
\label{sec:kinematics}

Given the promising results for the inelastic structure factors, we
calculate the expected recoil spectra for inelastic dark matter
scattering. The WIMP-nucleus scattering must conserve momentum and
energy:
\begin{eqnarray} 
{\bf q}_i &=& {\bf q}_f + {\bf q} \,, \\[1mm]
\frac{q_i^2}{2m_\chi} &=& \frac{q_f^2}{2m_\chi} + E_R + E^* \,,
\label{eq:EnergyEq} 
\end{eqnarray}
where ${\bf q}_i$ and ${\bf q}_f$ are the initial and final WIMP
momenta, ${\bf q}= {\bf q}_i - {\bf q}_f$ ($ = -{\bf p})$ is the
momentum transfer, $m_\chi$ is the WIMP mass, and $E^*$ the excitation
energy of the nucleus. The nuclear recoil energy is given by
\begin{equation} 
E_R = \frac{q^2}{2m_A} \,,
\label{eq:ER}
\end{equation}
with mass of the nucleus $m_A$. Eliminating $q_f$ in
Eq.~\eqref{eq:EnergyEq}, this leads to a quadratic equation for $q$,
\begin{equation}
q^2 - (2\mu v_i\cos\beta)q + 2\mu E^* = 0 \,,
\end{equation}
with the reduced mass $\mu=m_A m_\chi/(m_A+m_\chi)$, initial WIMP
velocity $v_i = q_i/m_\chi$, and where $\beta$ is the angle between
${\bf q_i}$ and ${\bf q}$. This equation has two solutions:
\begin{equation}
q_\pm = \mu v_i \cos\beta \left(1\pm\sqrt{1-\frac{2E^*}{\mu v_i^2 \cos^2\beta}} 
\right) \,.
\end{equation}
This leads to two constraints:
\begin{equation}
E^* \leqslant \frac{1}{2} \mu v_i^2 \cos^2\beta \leqslant 
\frac{1}{2} \mu v_i^2 \,,
\label{eq:E*up} 
\end{equation}
and 
\begin{equation}
v_i = \frac{1}{\cos\beta} \left(\frac{q}{2\mu}+\frac{E^*}{q}\right) \geqslant
\frac{q}{2\mu}+\frac{E^*}{q} = v_\text{min} \,,
\label{eq:vmin} 
\end{equation} 
where the allowed range is $\cos\beta \geqslant 0$. The upper limit in
Eq.~\eqref{eq:E*up} on the excitation energy shows that the inelastic
excitation is more sensitive to WIMPs with high velocities in the tail
of the dark matter halo distribution. Because $q_-$ ($q_+$) is
monotonically decreasing (increasing) with respect to $v_i\cos\beta$,
the absolute minimal (maximal) momentum transfer occurs at
$v_i\cos\beta=v_i$ for $\cos\beta=1$. This defines the minimal and
maximal recoil energies:
\begin{equation}
E_{R,\text{min}/\text{max}} = \frac{(\mu v_i)^2}{2m_A} 
\left(1 \mp \sqrt{1-\frac{2E^*}{\mu v_i^2 }}\,\right)^2 \,.
\label{eq:ERminmax}
\end{equation}
In the Earth's rest frame, the maximal $v_i$ is
given by $v_\text{esc}+v_{\text{Earth}}$, where $v_\text{esc}$ and $v_\text{\text{Earth}}$ denote
the galactic escape velocity and Earth's speed in the galaxy, respectively. We assume $v_\text{esc}=544$~km/s and a average $v_\text{\text{Earth}}=232$~km/s in subsequent calculations.  As $\frac{1}{2}\mu v_i^2 \to E^*$, the
domain of recoil energies over which the recoil spectrum is defined
shrinks, converging to the value $(\mu v_i)^2/2m_A = \mu E^*/m_A$.
Table~\ref{table:energydomain} gives the minimal and maximal recoil
energies for the two xenon isotopes and various WIMP masses.

\begin{table}[t]
\centering
\begin{tabular}{c c c c c}
\hline\hline
& \multicolumn{2}{c}{$^{129}$Xe} & \multicolumn{2}{c}{$^{131}$Xe} \\
Mass [GeV] & \, $E_{R,\mathrm{min}}$ & ${E_{R,\mathrm{max}}}$ & 
\, $E_{R,\mathrm{min}}$ & ${E_{R,\mathrm{max}}}$ \\ [0.5ex]
\hline
 10 & $-$ & $-$ & $-$ & $-$ \\
 25 & 1.5 & 31 & $-$ & $-$ \\
 50 & 1.2 & 110 & 6.8 & 81 \\
100 & 1.1 & 285 & 5.4 & 244 \\
250 & 1.1 & 659 & 4.9 & 601 \\
500 & 1.1 & 954 & 4.7 & 885 \\
\hline\hline
\end{tabular}
\caption{Minimal and maximal recoil energies, in keV, between which
inelastic scattering is allowed (see Eq.~\eqref{eq:ERminmax}) for the
two xenon isotopes and various WIMP masses.\label{table:energydomain}}
\end{table}

We calculate the nuclear recoil spectra for elastic and inelastic,
spin-dependent WIMP scattering off $^{129}$Xe and $^{131}$Xe following
Ref.~\cite{Smith:1988kw}:
\begin{eqnarray}
\frac{dR}{dE_R} &=& \frac{\sqrt{\pi}v_0}{2} \, \frac{R_0}{m_\chi m_A} \,
\frac{g(v_{\text{min}})}{E_0\,r} \nonumber \\ 
&&\times \frac{\sigma}{10^{-36}\,\mathrm{cm}^2} \,
\frac{\rho_0}{0.3\,\mathrm{GeVcm}^{-3}} \,
\frac{v_0}{220\,\mathrm{km\,s}^{-1}} \,,
\label{eq:Rate}
\end{eqnarray}
where the WIMP-nucleus cross section $\sigma$ is given by~\cite{Engel:1992bf}:
\begin{equation} 
\sigma = \frac{4}{3}\frac{\pi}{2J_i+1} 
\left(\frac{\mu}{\mu_{\text{nucleon}}}\right)^2 
S_A(q) \, \sigma_{\text{nucleon}} \,.
\end{equation}
$R_0=361$\,events/(kg\,d) is the total event rate per unit mass
for the Earth being at rest and an infinite
escape velocity, $E_0$ is the most probable kinetic energy of an
incident WIMP (given in terms of the the characteristic parameter of
the Maxwell-Boltzmann distribution $v_0$), $\rho_0$ is the local WIMP
density in our galaxy~\cite{Bovy:2012tw}, and $r$ is a kinematic
factor $r=4\mu/(m_A+m_\chi)$. $J_i$ is the nuclear spin in the
initial state, $\mu_{\text{nucleon}}$ is the WIMP-nucleon reduced
mass, and $\sigma_\text{nucleon}$ is the zero momentum transfer
cross-section for the nucleon. $g(v_{\text{min}})$ is the integral
containing information about the WIMP velocity distribution $f({\bf
v}+{\bf v}_{\text{Earth}})$~\cite{Mao:2013nda}:
\begin{equation}
g(v_{\text{min}}) = \int_{v_\text{min}}^\infty 
\frac{f({\bf v}+{\bf v}_{\text{Earth}})}{v} \, d^3{\bf v} \,,
\end{equation}
where the integral is from $v_{\text{min}}$, and the velocity
distribution $f({\bf v}+{\bf v}_{\text{Earth}})$ is truncated at
$v_{\text{esc}}$.

\begin{figure}[t]
\begin{center}
\includegraphics[width=0.975\columnwidth,clip=]{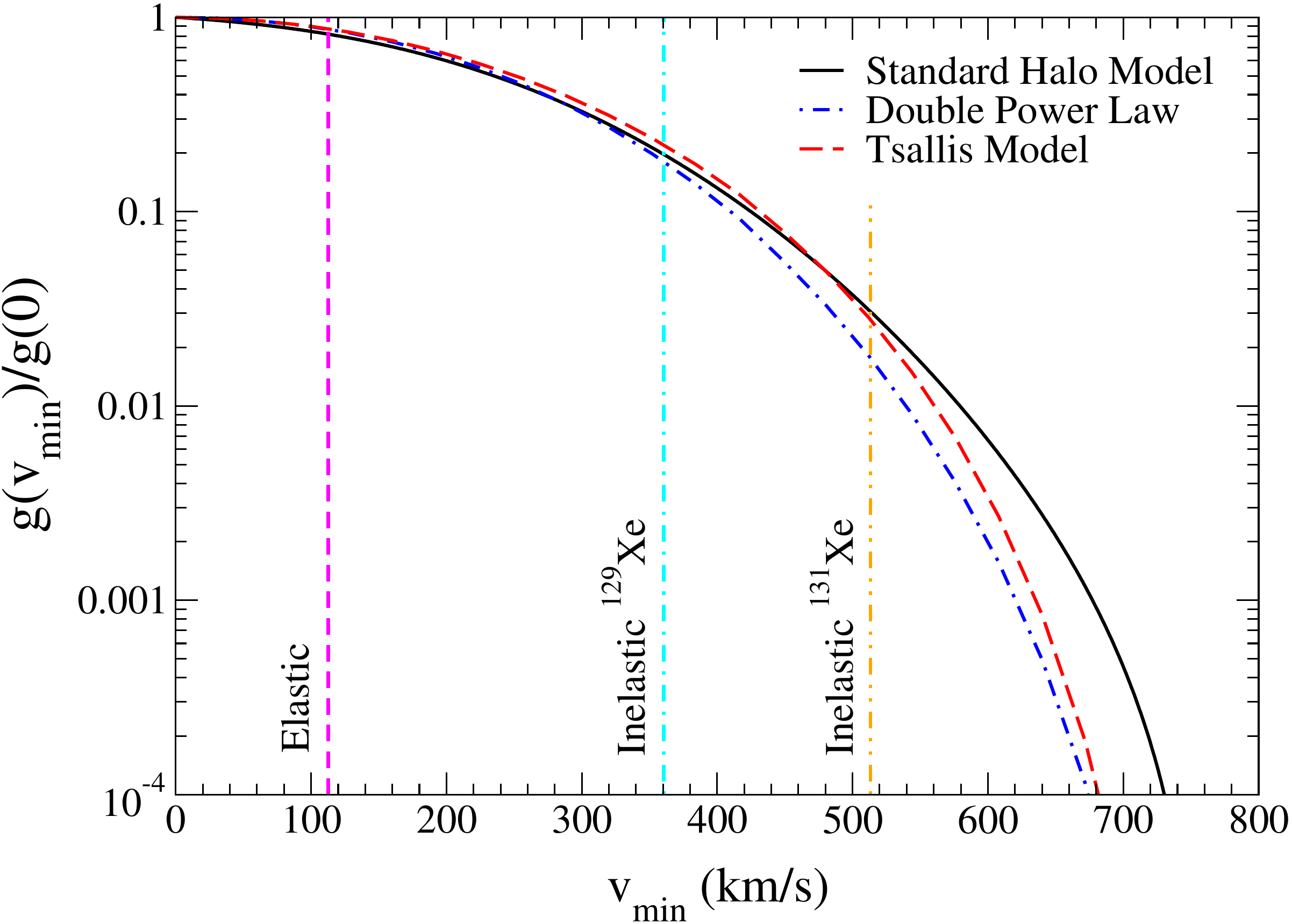}
\end{center}
\caption{(Color online) $g(v_{\text{min}})/g(0)$ as a function of $v_{\text{min}}$ for 
several WIMP velocity distributions. The lowest values of
$v_{\text{min}}$ are shown for elastic Xe (magenta), inelastic
$^{129}$Xe (cyan), and inelastic $^{131}$Xe (yellow) scattering. In
the elastic case, $v_{\text{min}}$ is calculated at XENON100's nuclear
recoil energy threshold of $\sim$\,7\,keV (neglecting the difference in
the nuclear mass number $A$)~\cite{Aprile:2012nq}.\label{fig:VelDist}}
\end{figure}

Figure~\ref{fig:VelDist} shows the normalized velocity integral, $g(v_{\text{min}})/g(0)$, for three WIMP
velocity distributions ~\cite{Frandsen:2011gi}: 
the Standard Halo
Model~\cite{Frandsen:2011gi}, the Double Power Law
profile~\cite{Navarro:1995iw}, and the Tsallis
model~\cite{Ling:2009eh, Vergados:2007nc}. The different types of
interactions are sensitive to different velocities. While for elastic
scatters, the lowest $v_{\text{min}}$ is determined by a detector's
energy threshold, in the inelastic case the lowest $v_{\text{min}}$
occurs at $\sqrt{2 E^*/\mu}$ (or recoil energy $E_R= \mu E^*/m_A$,
see the discussion above).

\begin{figure}[t]
\begin{center}
\includegraphics[width=0.975\columnwidth,clip=]{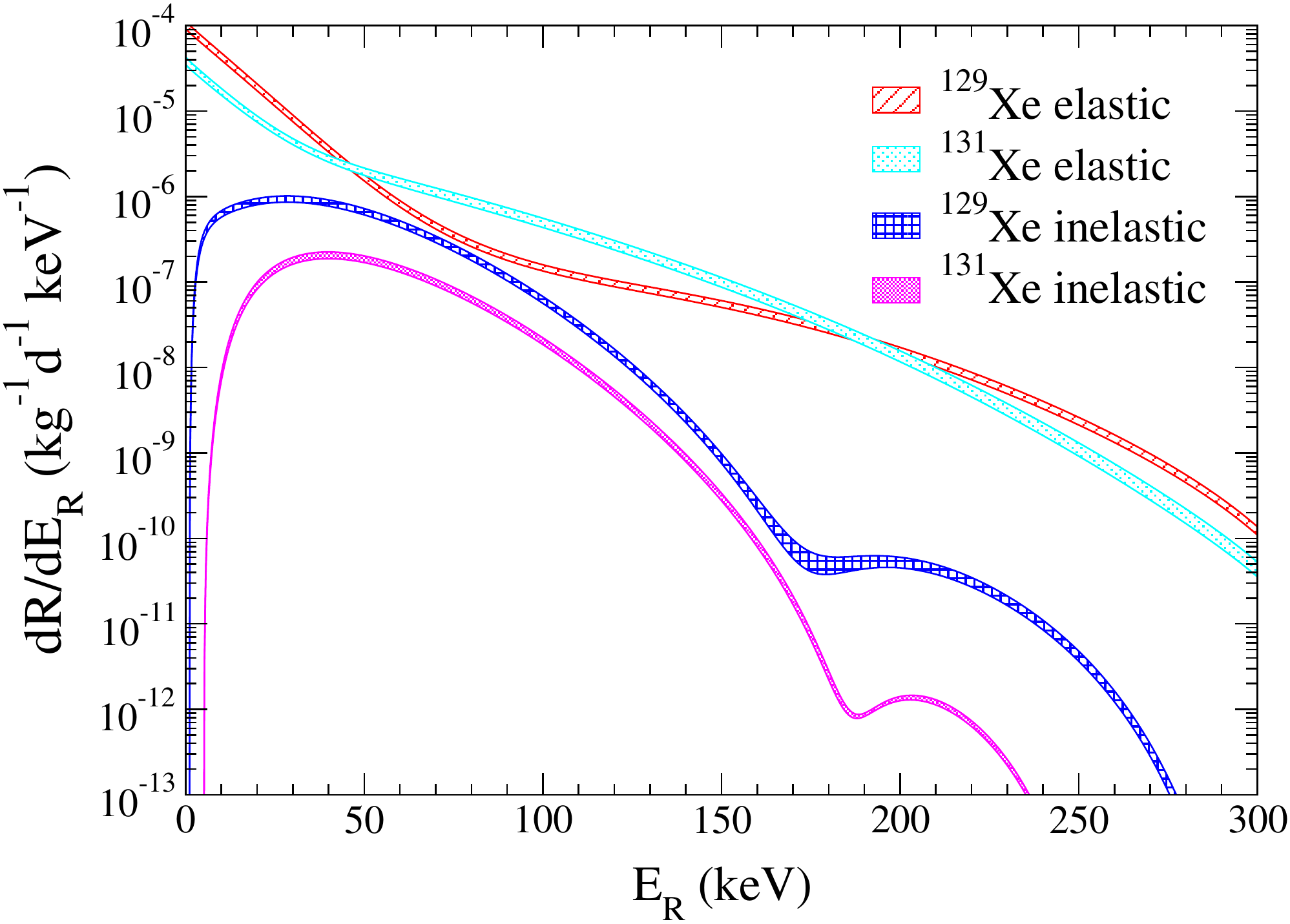}
\end{center}
\caption{(Color online) Differential nuclear recoil spectra $dR/dE_R$ as a function
of recoil energy $E_R$ for scattering off $^{129}$Xe and $^{131}$Xe,
assuming ``neutron-only'' couplings of a 100\,GeV WIMP with a
WIMP-nucleon cross section
$\sigma_{\text{nucleon}}=10^{-40}$\,cm$^2$. Both elastic and inelastic
recoil spectra are shown for comparison, including chiral 1b+2b
currents, where the bands include the uncertainties due to
WIMP-nucleon currents.\label{fig:recoilspectra}}
\end{figure}

Figure~\ref{fig:recoilspectra} shows the differential recoil spectra
for scattering off $^{129}$Xe and $^{131}$Xe for ``neutron-only''
couplings and for the structure factors that include chiral 1b+2b
currents. The elastic structure factors are taken from
Ref.~\cite{Klos}. The widths of the bands reflect the theoretical
uncertainties in the WIMP-nucleon currents. The elastic,
spin-dependent spectra are shown for comparison. For illustration, we
show all spectra here and in the following over a wide range of
expected rates but note that, for the interpretation of future
experimental results, only the highest-rate parts will be relevant. As
expected, the inelastic spectra fall to zero at the bounds of the
recoil energy, and the domain for $^{131}$Xe is smaller because of the
higher excitation energy compared to $^{129}$Xe.

\section{Signatures of inelastic dark matter scattering}
\label{sec:signatures}

Dark matter detectors that use liquid xenon can be expected among those
most sensitivite to inelastic, spin-dependent WIMP
scattering~\cite{Aprile:2013doa}. In these detectors, the nuclear recoil
and subsequent gamma emission will occur at the same space-time
coordinates: the half-lives of the lowest excited states in $^{129}$Xe
and $^{131}$Xe are 0.97\,ns and 0.48\,ns, respectively, and the mean
free paths of gammas with energies of 39.6\,keV and 80.2\,keV in
liquid xenon are $\sim$\,0.15\,mm and 0.92\,mm, whereas experimental
resolutions are typically of order 10\,ns and 3\,mm~\cite{Aprile:2011dd}.
Experiments can therefore search for a
total energy deposition,
\begin{equation}
E_{\mathrm{vis}} = f(E_R) \times E_R + E^* \,,
\end{equation}
in the detector, where $E_{\mathrm{vis}}$ is the observed energy and
$f(E_R)$ is an energy-dependent quenching factor for nuclear
recoils~\cite{Smith:1988kw}. Only a fraction $f(E_R)$ of the nuclear
recoil energy will be transferred to electronic excitations. These can
be observed as scintillation light in a single-phase detector or as
prompt scintillation and delayed charge signals in liquid xenon time
projection chambers~\cite{Chepel:2012sj,Aprile:1900zz}; the rest is
transferred to heat and remains undetected.

\begin{figure}[t]
\begin{center}
\includegraphics[width=0.975\columnwidth,clip=]{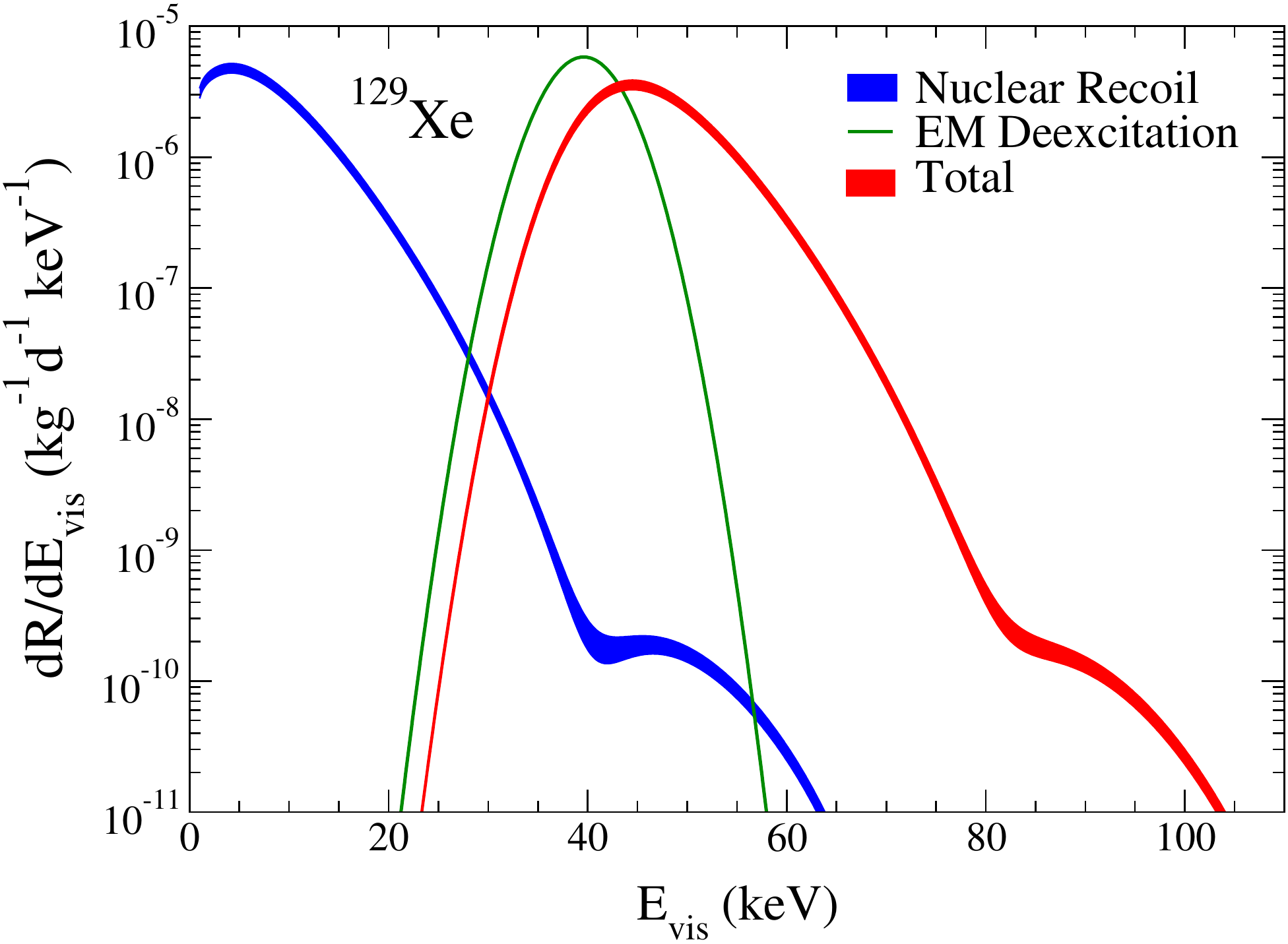}
\includegraphics[width=0.975\columnwidth,clip=]{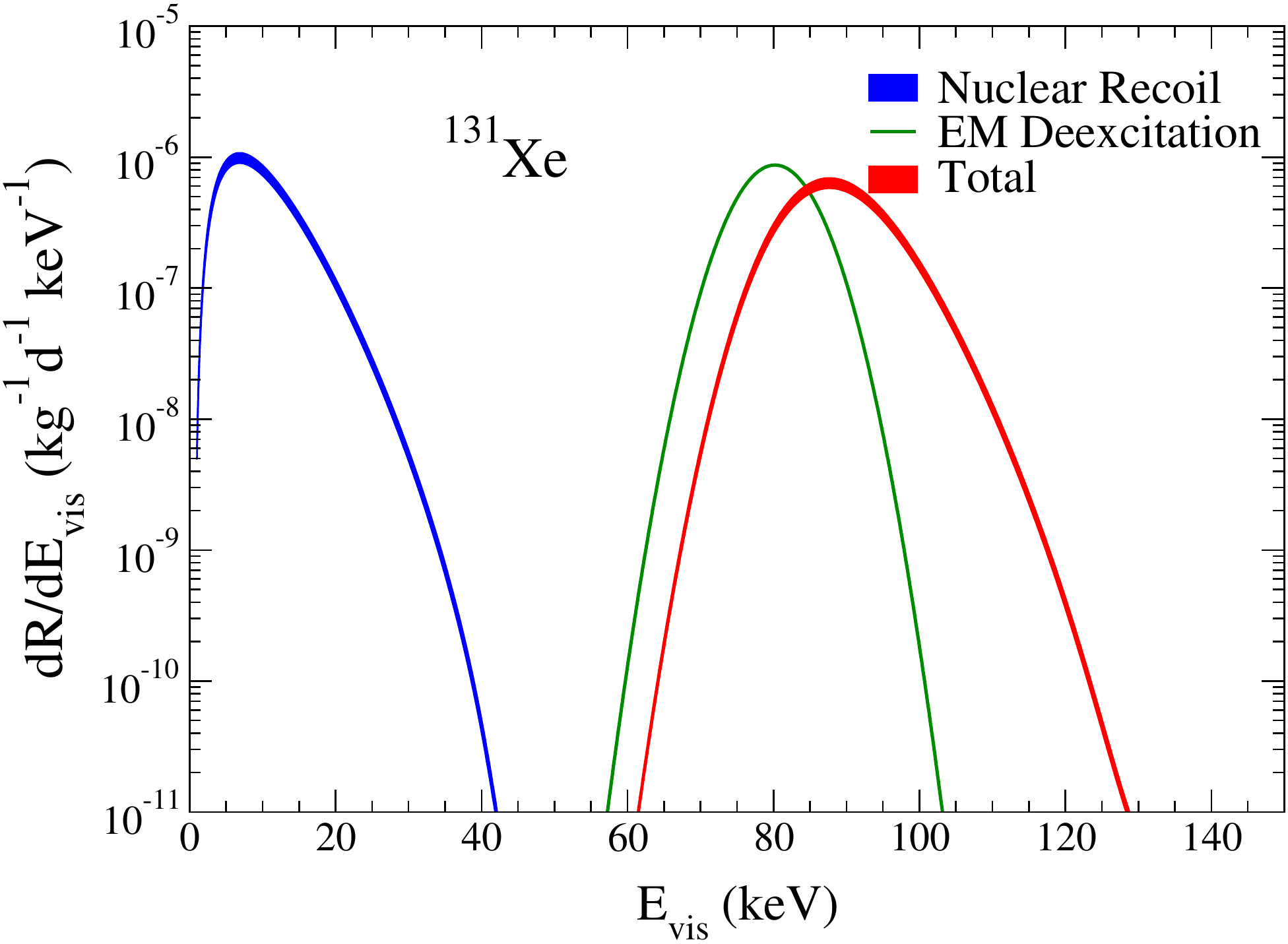}
\end{center}
\caption{(Color online) Differential energy spectra as a function of $E_\text{vis}$
for inelastic scattering off $^{129}$Xe (top) and $^{131}$Xe
(bottom). Shown is the nuclear recoil spectrum (blue), the
de-excitation gamma (green), and their sum (red), assuming realistic
detector resolution and quenching of the nuclear recoil signal. As in
Figure~\ref{fig:recoilspectra}, results are shown with chiral 1b+2b
currents, where the bands include the uncertainties due to
WIMP-nucleon currents, and we have assumed ``neutron-only'' couplings
of a 100\,GeV WIMP with
$\sigma_{\text{nucleon}}=10^{-40}$\,cm$^2$.\label{fig:diffspec}}
\end{figure}

Figure~\ref{fig:diffspec} shows the differential nuclear recoil
spectra for inelastic scattering off $^{129}$Xe and $^{131}$Xe, the
differential energy spectra of the de-excitation gammas, and the sum
of these two contributions. The energy scale is based on the total
number of quanta detected in a liquid xenon dark matter
experiment. For nuclear recoils, we assume the Lindhard
theory~\cite{Lindhard} with a conservative choice of the
proportionality constant between the electronic stopping power and the
velocity of the recoiling xenon atom, $k=0.110$, as shown in Figure~1 of
Ref.~\cite{Sorensen:2011bd}. For the gamma lines, we assume an energy
resolution as obtained in the XENON100 detector at these energies in
inelastic neutron-xenon scatters~\cite{Aprile:2011dd}, using a linear
combination of the primary scintillation and proportional
scintillation signals~\cite{Aprile:2007qd}: $\sigma/E = 9\%$ at
40\,keV and $\sigma/E = 6.5\%$ at 80\,keV.

\begin{figure}[t]
\begin{center}
\includegraphics[width=0.975\columnwidth,clip=]{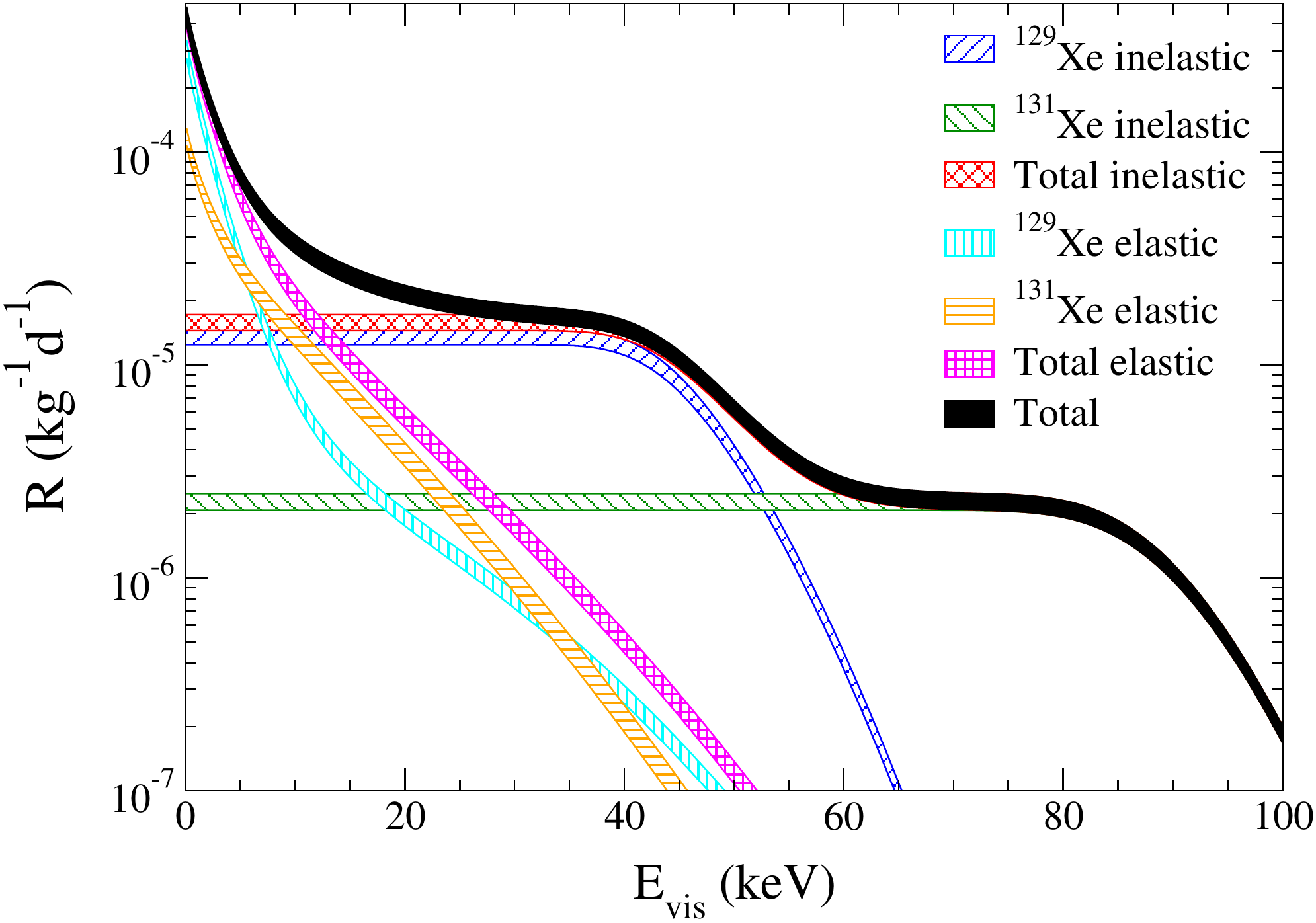}
\end{center}
\caption{(Color online) Integrated energy spectra of xenon for elastic and inelastic,
spin-dependent scattering for ``neutron-only'' couplings and a
100\,GeV WIMP with $\sigma_{\text{nucleon}}=10^{-40}$\,cm$^2$. The
differential spectra are integrated from a given threshold value
$E_{\mathrm{vis}}$ to infinity. The inelastic contributions dominate
over the elastic ones for moderate energy thresholds.  As in
Figure~\ref{fig:recoilspectra}, results are shown with chiral 1b+2b
currents, where the bands include the uncertainties due to
WIMP-nucleon currents.\label{fig:intspecxe100}}
\end{figure}

\begin{figure}[t]
\begin{center}
\includegraphics[width=0.975\columnwidth,clip=]{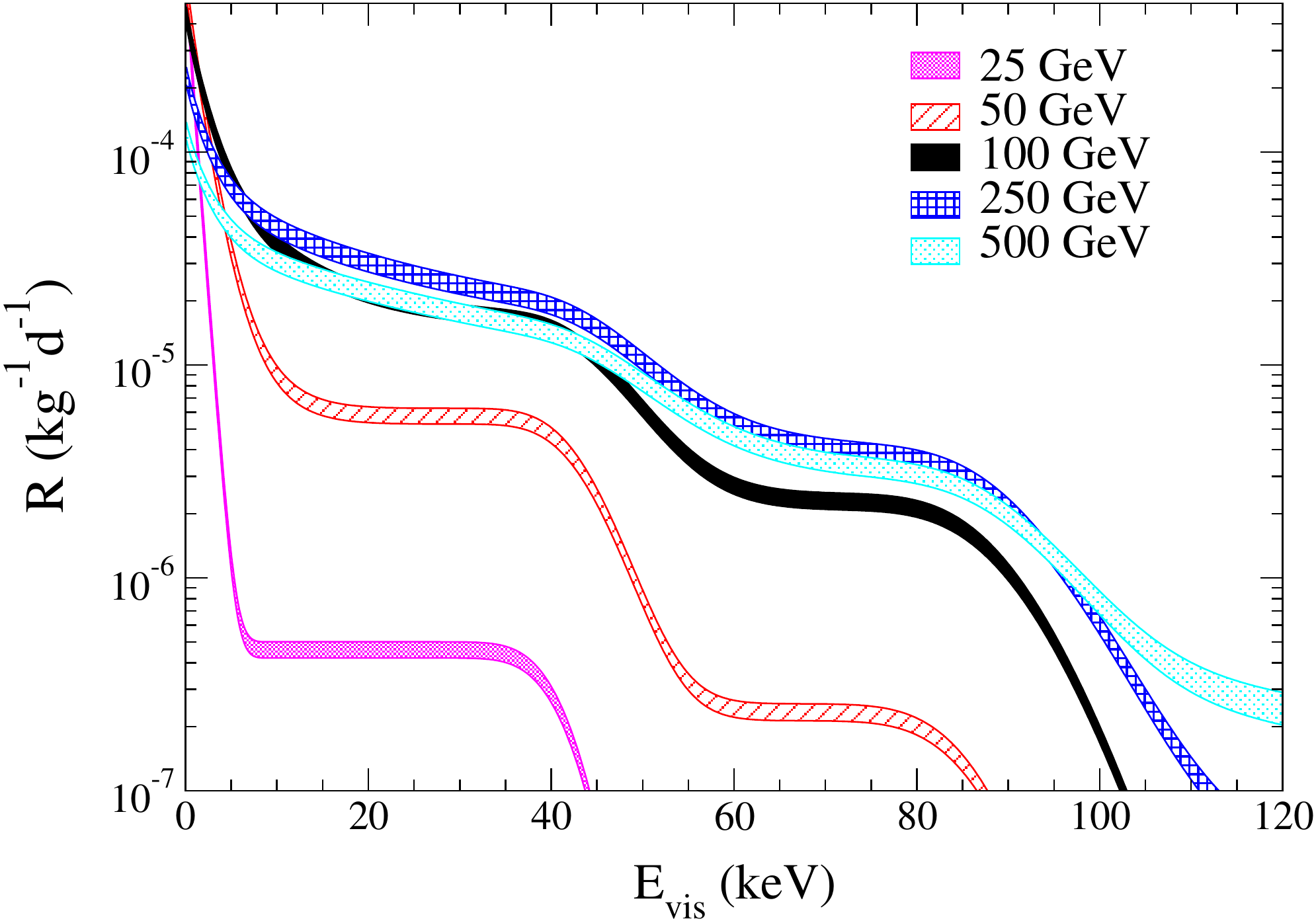}
\end{center}
\caption{(Color online) Integrated total energy spectra of xenon as in 
Figure~\ref{fig:intspecxe100} for various WIMP masses from
$25-500$\,GeV. The solid black curve is for a 100\,GeV WIMP as in
Figure~\ref{fig:intspecxe100}. In all cases, minimum energies in
$E_{\mathrm{vis}}$ exist, above which the inelastic channel dominates
the elastic one. The spectrum for a 500\,GeV WIMP is lower than the
for 250\,GeV, because the particle number density in the galaxy
decreases with increasing mass.\label{fig:intspecxe25-500}}
\end{figure}

Due to the prompt gamma from the nuclear de-excitation, the region of
interest for the dark matter search is shifted to higher energies
compared to elastic scattering. Thus, experiments can search in their
observed differential energy spectra for clear structures around
40\,keV ($^{129}$Xe) and 80\,keV ($^{131}$Xe), although the higher
energy region is kinematically suppressed. This suppression is
particularly pronounced for WIMP masses below $\sim$\,100\,GeV (see
Figure~\ref{fig:intspecxe25-500}).

In near-future detectors, observed event rates are expected to be
small. While a precise measurement of the differential energy spectrum
will thus be statistically prohibited, integrated rates can provide
valuable insight. Figure~\ref{fig:intspecxe100} shows the integrated
spectra for elastic and inelastic scattering of a 100\,GeV WIMP as
well as their sum, while Figure~\ref{fig:intspecxe25-500} displays
these for a range of WIMP masses.  These spectra are shown for natural xenon, thus the isotopic contributions of each isotope,
given by Eq. \eqref{eq:Rate}, are weighted by the aforementioned abundances.  The inelastic channel is favored
over the elastic one at momentum transfers above 
36\,MeV, 45\,MeV, 54\,MeV, 68\,MeV and 76\,MeV 
(corresponding to nuclear recoil energies of 
5\,keV, 8\,keV, 12\,keV, 19\,keV  and 24\,keV) for WIMP masses of 
25, 50, 100, 250 and 500\,GeV,
respectively. Should both elastic and inelastic signatures be observed
by a current or future xenon experiment, it would offer a strong case
for the spin-dependent nature of WIMP-nucleon interactions. On the
other hand, if signal-like events are observed with no excess signal
in the spin-dependent inelastic channels, this would indicate a
spin-independent nature of WIMP-nucleon interactions.

\section{Implications for dark matter searches and conclusions}
\label{sec:implications}

Current xenon-based dark matter experiments are either single-phase
(liquid) detectors or dual-phase (liquid/gas) time projection chambers
(TPCs)~\cite{Chepel:2012sj,Aprile:1900zz}.  In single-phase detectors,
a large liquid xenon volume is instrumented with photodetectors, and
the prompt scintillation light signal is observed.  In a TPC, the
prompt light signal is observed with two arrays of photosensors, and,
in addition, free electrons are drifted away from the interaction site
toward the vapor phase above the liquid and detected via an amplified,
proportional scintillation light signal. In both cases, background
reduction is achieved by self-shielding and fiducialization, namely
the selection of an inner, low-background liquid xenon volume based on
the reconstructed vertex of each event. For TPCs, an additional tool
to distinguish background from signal events is the charge-to-light
ratio~\cite{Aprile:2006kx}, which depends on the electronic stopping
power d$E/$d$x$, and hence on the type of particle interaction.

Operational liquid xenon dark matter experiments such as
XENON100~\cite{Aprile:2011dd} and XMASS~\cite{Abe:2013tc} have reached
overall background levels of
$\sim$\,0.01\,events/(kg\,d\,keV)~\cite{Aprile:2011vb} and
$\sim$\,0.7\,events/(kg\,d\,keV)~\cite{Abe:2012ut}, respectively, in
the energy region $<$\,200\,keV.  This implies sensitivities to
inelastic, spin-dependent WIMP-nucleon cross sections of
3.8$\times$10$^{-36}$\,cm$^{2}$ (XENON100) and
1.3$\times$10$^{-33}$\,cm$^{2}$ (XMASS) for a WIMP with mass
$130$\,GeV.  These sensitivities are four (XENON100) and seven (XMASS)
orders of magnitude worse than the current best limit for
spin-dependent interactions from XENON100 at this WIMP mass, but could
be improved by a dedicated analysis. A temporal or spatial coincidence
between the nuclear recoil and the de-excitation gamma cannot be
employed to further suppress the background for aforementioned 
reasons related to detector resolution. However, the observed background spectra are
flat in this energy range, and no peak-like structures are expected
from internal (such as $^{222}$Rn and $^{85}$Kr) or external
sources. Fast neutrons can in principle inelastically scatter off
xenon nuclei, and in fact, interactions from $^{241}$AmBe neutrons or
similar sources are used as calibration sources for both elastic and
inelastic interactions~\cite{Aprile:2011dd}. However, current dark
matter experiments are designed such that a background level of
$\ll$\,1 elastic nuclear recoil per given exposure is expected from
neutrons, implying an even lower rate for inelastic neutron
scatters. Hence, a dedicated analysis can search for a peak-like
enhancement in an otherwise flat background spectrum.  Additionally,
two-phase detectors can discriminate between the purely electronic
recoil background and the WIMP-induced signal, which contains an
additional nuclear recoil component based on the charge-to-light ratio.

The LUX experiment~\cite{Akerib:2012ys}, which started a first physics
run in 2013, is expected to reach a background level that is a factor
10 below the one of XENON100. Future experiments such as
XMASS-5t~\cite{xmass5t}, XENON1T~\cite{Aprile:2012zx},
LZ~\cite{Malling:2011va}, and DARWIN~\cite{Baudis:2012bc,Schumann:2011ts}
are to lower their backgrounds by yet another factor of $100-1000$,
compared to XENON100. This reduction can be expected to continue until 
the overall electronic-recoil background will be dominated by irreducible
neutrino-electron scatters from solar neutrinos at the level of
8$\times$10$^{-6}$\,\,events/(kg\,d\,keV)~\cite{Baudis:2012bc}, at which 
point further improvements in sensitivity to this interaction channel
will be inhibited.

A particular experiment, using the same detector, target, and isotopic
abundances, can measure both the elastic and inelastic recoil spectra.
Because, for a given halo model, the ratio of the cross sections for
elastic and inelastic scattering is fixed, these measured spectra
would provide insight into the characteristics of the dark matter
halo. For instance, the relative contributions of the elastic and
inelastic scattering to the spin-dependent channel may be compared in
order to inspect the WIMP velocity distribution. The fraction of the
event rate that each channel contributes is necessarily connected to
the integral of the velocity distribution given in Eq.~\eqref{eq:Rate}
and shown in Figure~\ref{fig:VelDist}, such that any deviation from
the expected contributions would reflect a deviation in the expected
WIMP velocity distribution.

The mass of the dark matter particle is uniquely connected to both the
recoil energy domain and to the point at which the inelastic channel
dominates the elastic one, thereby providing two methods by which to
use the observed spectra to probe the WIMP mass.
Table~\ref{table:crossing} lists the energies at which the inelastic
spectrum starts to dominate the elastic one for the WIMP masses
considered in Sect.~\ref{sec:signatures}. Due to the kinematic
constraints discussed in Sect.~\ref{sec:kinematics}, inelastic
scattering is not possible for $^{129}$Xe at WIMP masses below 14\,GeV
or for $^{131}$Xe at masses below 31\,GeV.

\begin{table}[t]
\centering
\begin{tabular}{c c c c}
\hline\hline
Mass [GeV] & \: $^{129}$Xe \: & \: $^{131}$Xe \: & \: Total \\\hline
10 & $-$ & $-$ & \: $-$ \\
25 & 5 & $-$ & \: 5 \\
50 & 7 & 17 & \: 9 \\
100 & 7 & 24 & \: 12 \\
250 & 9 & 32 & \: 19 \\
500 & 11 & 35 & \: 24 \\
\hline\hline
\end{tabular}
\caption{Minimum energy $E_\text{vis}$ in keV above which the observed
inelastic spectrum for $^{129}$Xe, $^{131}$Xe and for the total
spectrum starts to dominate the elastic one for various WIMP
masses.\label{table:crossing}}
\end{table}

Once a potential dark matter signal is observed in a xenon-based
experiment, one could use a target enriched or depleted in $^{129}$Xe
and $^{131}$Xe. Thus, one would control the ratio of odd-even isotopes
to enhance or deplete certain types of interactions. While the
observed rate in the spin-independent channel would remain roughly
unchanged, the rate in the spin-dependent channel, both elastic and
inelastic, would increase or decrease with the fraction of odd xenon
isotopes in the detector.

Inelastic scattering of dark matter can also be studied in other
isotopes considered for searches of spin-dependent WIMP
scattering. The nucleus $^{73}$Ge would be especially
promising~\cite{Holmlund:2004rv} as it has a very low-lying state at
13\,keV, plus two other low-lying excited states at 67 and 69\,keV. In
addition, $^{127}$I~\cite{Vergados:2013raa} has a low-lying state at
58\,keV. The first excited states of lighter isotopes relevant to
experimental searches ($^{19}$F and $^{23}$Na) have higher excitation
energies $>100$\,keV. Combined with the smaller reduced WIMP-nucleus
mass, these lighter nuclei are therefore not sensitive to inelastic
scattering.

In conclusion, we have shown that for spin-dependent WIMP-nucleus
interactions, inelastic scattering dominates over elastic scattering
if the momentum transfer is above a value that is still kinematically
accessible for galactic dark matter to scatter off a terrestrial xenon
target. To this end, we performed detailed calculations of the
inelastic, spin-dependent nuclear structure factors for $^{129}$Xe and
$^{131}$Xe, which include estimates of the theoretical uncertainties
in WIMP-nucleon currents. If the observed energy in a given detector
is limited to a value above $4-16$\,keV, depending on the WIMP mass,
the sensitivity to spin-dependent WIMP-nucleon interactions will be
given by the inelastic channel alone. This can have important
consequences for the appropriate analysis strategy. In addition, it is
conceivable that detectors that are optimized for a higher energy
threshold but lower radioactive background levels~\cite{Auger:2012ar}
can obtain competitive spin-dependent WIMP sensitivity using this
channel alone, provided their target contains an isotope with an
exited state that is kinematically accessible. Finally, this
additional detection channel can provide useful information to
disentangle a potential nuclear recoil signal from other nuclear
recoil backgrounds. If signal events are observed in the elastic
channel, simultaneous detection of signal events in the inelastic
channel will point towards a spin-dependent nature of the
interaction. With sufficient statistics, data from a single detector
can be analyzed to extract information about the particle mass and the
dark matter halo in a fully complementary way, breaking degeneracies
otherwise inherent to analyses focusing on the elastic channel alone.

\section*{Acknowledgements}

This work was supported by the SNF through Grant 200020-138225, the
Helmholtz Association through the Helmholtz Alliance Program, contract
HA216/EMMI ``Extremes of Density and Temperature: Cosmic Matter in the
Laboratory'', the DFG through Grant SFB 634, the NSF under Grant
No.~1206061, and the Purdue Research Foundation. We would like to thank the 
referee for very useful comments.


\onecolumngrid

\begin{table}[t]
\centering {\bf Appendix}
\caption{Fits to the isoscalar/isovector structure factors $S_{00}$, 
$S_{11}$ and $S_{01}$ as well as ``proton-only" and ``neutron-only"
structure factors $S_{p}$ and $S_{n}$ for spin-dependent WIMP elastic
scattering off $^{129}$Xe and $^{131}$Xe nuclei from Ref.~[15],
including 1b and 2b currents as in Fig. 2. The upper and lower limits
from the theoretical error band were used for the fit.  The fitting
function of the dimensionless variable $u = p^2 b^2/2$ is $S_{ij}(u) =
e^{-u} \sum_{n=0}^9 c_{ij,n} u^n$. The rows give the coefficients
$c_{ij,n}$ of the $u^n$ terms in the polynomial.}
\begin{center}
\begin{tabular*}{0.775\textwidth}{c||c|c|c|c|c}
\hline
\multicolumn{6}{c}{$^{129}$Xe elastic} \\
\multicolumn{6}{c}{$u=p^2b^2/2 \,, \: b=2.2853 \, {\rm fm}$} \\
\hline
$e^{-u}\times$ & $S_{00}$ & $S_{11}$ (1b+2b min) & $S_{11}$ (1b+2b max) & 
$S_{01}$ (1b+2b min) & $S_{01}$ (1b+2b max) \\
\hline
$1$ & $0.0547144$ & $0.0221559$ & $0.0357742$ & $-0.0885644$ & $-0.0696691 $\\
$u$ & $-0.146407$ & $-0.0656100$ & $-0.107895$ & $0.254049$ & $0.197380 $\\
$u^2$ & $0.180603$ & $0.0863920$ & $0.145055$ & $-0.332322$ & $-0.254839 $\\
$u^3$ & $-0.125526$ & $-0.0631729$ & $-0.108549$ & $0.244981$ & $0.185896 $\\
$u^4$ & $0.0521484$ & $0.0278792$ & $0.0490401$ & $-0.109298$ & $-0.0825294 $\\
$u^5$ & $-0.0126363$ & $-0.00756661$ & $-0.0136169$ & $0.0296705$ & $0.0224322 $\\
$u^6$ & $0.00176284$ & $0.00126767$ & $0.00233283$ & $-0.00492657$ & $-0.00375109 $\\
$u^7$ & $-1.32501 \times 10^{-4}$ & $-1.27755 \times 10^{-4}$ & $-2.39926 \times 10^{-4}$ & $4.88467 \times 10^{-4}$ & $3.77179 \times 10^{-4} $\\
$u^8$ & $4.23423 \times 10^{-6}$ & $7.10322 \times 10^{-6}$ & $1.35553 \times 10^{-5}$ & $-2.65022 \times 10^{-5}$ & $-2.09510 \times 10^{-5} $\\
$u^9$ & $-1.68052 \times 10^{-9}$ & $-1.67272 \times 10^{-7}$ & $-3.21404 \times 10^{-7}$ & $5.98909 \times 10^{-7}$ & $4.92362 \times 10^{-7} $\\
\hline
$e^{-u}\times$ & & $S_{p}$ (1b+2b min) & $S_{p}$ (1b+2b max)& 
$S_{n}$ (1b+2b min) & $S_{n}$ (1b+2b max) \\
\hline
$1$&&$ 0.00196369$ & $0.00715281$ & $0.146535$ & $0.179056 $\\
$u$&&$ -0.00119154$ & $-0.0134790$ & $-0.409290$ & $-0.508334 $\\
$u^2$&&$ -0.00324210$ & $0.00788823$ & $0.521423$ & $0.657560 $\\
$u^3$&&$ 0.00622602$ & $0.00311153$ & $-0.374011$ & $-0.477988 $\\
$u^4$&&$ -0.00496653$ & $-0.00653771$ & $0.162155$ & $0.209437 $\\
$u^5$&&$ 0.00224469$ & $0.00375478$ & $-0.0424842$ & $-0.0554186 $\\
$u^6$&&$ -5.74412 \times 10^{-4}$ & $-0.00105558$ & $0.00674911$ & $0.00889251 $\\
$u^7$&&$ 8.31313 \times 10^{-5}$ & $1.59440 \times 10^{-4}$ & $-6.33434 \times 10^{-4}$ & $-8.42977 \times 10^{-4} $\\
$u^8$&&$ -6.41114 \times 10^{-6}$ & $-1.25055 \times 10^{-5}$ & $3.20266 \times 10^{-5}$ & $4.30517 \times 10^{-5} $\\
$u^9$&&$ 2.07744 \times 10^{-7}$ & $4.04987 \times 10^{-7}$ & $-6.54245 \times 10^{-7}$ & $-8.88774 \times 10^{-7} $\\
\hline
\hline
\multicolumn{6}{c}{$^{131}$Xe elastic} \\
\multicolumn{6}{c}{$u=p^2b^2/2 \,, \: b=2.2905 \, {\rm fm}$} \\
\hline 
$e^{-u}\times$ & $S_{00}$ & $S_{11}$ (1b+2b min) & $S_{11}$ (1b+2b max) &
$S_{01}$ (1b+2b min) & $S_{01}$ (1b+2b max) \\
\hline
$1$ & $0.0417857$ & $0.0167361$ & $0.0271052$ & $-0.0675438$ & $-0.0529487 $\\
$u$ & $-0.111132$ & $-0.0472853$ & $-0.0812985$ & $0.195710$ & $0.146987 $\\
$u^2$ & $0.171306$ & $0.0684924$ & $0.122960$ & $-0.306688$ & $-0.225003 $\\
$u^3$ & $-0.132481$ & $-0.0514413$ & $-0.0940491$ & $0.243678$ & $0.179499 $\\
$u^4$ & $0.0630161$ & $0.0237858$ & $0.0439746$ & $-0.118395$ & $-0.0888278 $\\
$u^5$ & $-0.0177684$ & $-0.00692778$ & $-0.0128013$ & $0.0351428$ & $0.0271514 $\\
$u^6$ & $0.00282192$ & $0.00124370$ & $0.00227407$ & $-0.00622577$ & $-0.00499280 $\\
$u^7$ & $-2.32247 \times 10^{-4}$ & $-1.31617 \times 10^{-4}$ & $-2.35642 \times 10^{-4}$ & $6.31685 \times 10^{-4}$ & $5.31148 \times 10^{-4} $\\
$u^8$ & $7.81471 \times 10^{-6}$ & $7.46669 \times 10^{-6}$ & $1.28691 \times 10^{-5}$ & $-3.33272 \times 10^{-5}$ & $-2.99162 \times 10^{-5} $\\
$u^9$ & $1.25984 \times 10^{-9}$ & $-1.73484 \times 10^{-7}$ & $-2.77011 \times 10^{-7}$ & $6.82500 \times 10^{-7}$ & $6.81902 \times 10^{-7} $\\
\hline
$e^{-u}\times$ & & $S_{p}$ (1b+2b min) & $S_{p}$ (1b+2b max)& 
$S_{n}$ (1b+2b min) & $S_{n}$ (1b+2b max) \\
\hline
$1$&&$ 0.00159352$ & $0.00529643$ & $0.111627$ & $0.136735 $\\
$u$&&$ -0.00207344$ & $-0.00528808$ & $-0.308602$ & $-0.393930 $\\
$u^2$&&$ 0.00567412$ & $-0.00627452$ & $0.474842$ & $0.617924 $\\
$u^3$&&$ -0.00605643$ & $0.0227436$ & $-0.375201$ & $-0.488443 $\\
$u^4$&&$ 0.00337794$ & $-0.0192229$ & $0.182382$ & $0.234645 $\\
$u^5$&&$ -6.88135 \times 10^{-4}$ & $0.00844826$ & $-0.0539711$ & $-0.0681357 $\\
$u^6$&&$ -3.42717 \times 10^{-5}$ & $-0.00212755$ & $0.00944180$ & $0.0116393 $\\
$u^7$&&$ 3.13222 \times 10^{-5}$ & $3.03972 \times 10^{-4}$ & $-9.34456 \times 10^{-4}$ & $-0.00111487 $\\
$u^8$&&$ -4.02617 \times 10^{-6}$ & $-2.27893 \times 10^{-5}$ & $4.73386 \times 10^{-5}$ & $5.34878 \times 10^{-5} $\\
$u^9$&&$ 1.72711 \times 10^{-7}$ & $7.05661 \times 10^{-7}$ & $-9.01514 \times 10^{-7}$ & $-9.03594 \times 10^{-7} $\\
\hline
\end{tabular*}
\end{center}
\end{table}

\newpage

\onecolumngrid

\begin{table}[t]
\caption{Fits to the isoscalar/isovector structure factors $S_{00}$, 
$S_{11}$ and $S_{01}$ as well as ``proton-only" and ``neutron-only" 
structure factors $S_{p}$ and $S_{n}$ for spin-dependent WIMP
inelastic scattering off $^{129}$Xe nuclei, including 1b and 2b
currents as in Fig. 1. The upper and lower limits from the theoretical
error band were used for the fit.  The fitting function of the
dimensionless variable $u = p^2 b^2/2$ is $S_{ij}(u) = e^{-u}
\sum_{n=0}^{13} c_{ij,n} u^n$. The rows give the coefficients 
$c_{ij,n}$ of the $u^n$ terms in the polynomial.}
\begin{center}
\begin{tabular*}{0.795\textwidth}{c||c|c|c|c|c}
\hline
\multicolumn{6}{c}{$^{129}$Xe inelastic } \\
\multicolumn{6}{c}{$u=p^2b^2/2 \,, \: b=2.2853 \, {\rm fm}$} \\
\hline
$e^{-u}\times$ & $S_{00}$ & $S_{11}$ (1b+2b min) & $S_{11}$ (1b+2b max) & 
$S_{01}$ (1b+2b min) & $S_{01}$ (1b+2b max) \\
\hline
$1$ & $0.00245755$ & $7.84441 \times 10^{-4}$ & $0.00126462$ & $-0.00352622$ & $-0.00277704 $\\
$u$ & $-0.00643918$ & $-0.00236681$ & $-0.00374626$ & $0.00984723$ & $0.00784581 $\\
$u^2$ & $0.0402663$ & $0.0178893$ & $0.0257771$ & $-0.0637666$ & $-0.0538274 $\\
$u^3$ & $-0.0499158$ & $-0.0275022$ & $-0.0363040$ & $0.0825757$ & $0.0752627 $\\
$u^4$ & $0.0268070$ & $0.0203782$ & $0.0228147$ & $-0.0449151$ & $-0.0495933 $\\
$u^5$ & $-0.00753795$ & $-0.00946420$ & $-0.00745241$ & $0.0102210$ & $0.0203582 $\\
$u^6$ & $0.00116620$ & $0.00318682$ & $0.00101110$ & $9.85679 \times 10^{-4}$ & $-0.00633059 $\\
$u^7$ & $-9.84820 \times 10^{-5}$ & $-8.40630 \times 10^{-4}$ & $1.70578 \times 10^{-4}$ & $-0.00142386$ & $0.00168005 $\\
$u^8$ & $5.13618 \times 10^{-6}$ & $1.74690 \times 10^{-4}$ & $-1.14316 \times 10^{-4}$ & $4.91504 \times 10^{-4}$ & $-3.72182 \times 10^{-4} $\\
$u^9$ & $-4.08657 \times 10^{-7}$ & $-2.73828 \times 10^{-5}$ & $2.72172 \times 10^{-5}$ & $-9.94860 \times 10^{-5}$ & $6.25058 \times 10^{-5} $\\
$u^{10}$ & $4.84340 \times 10^{-8}$ & $3.04920 \times 10^{-6}$ & $-3.79530 \times 10^{-6}$ & $1.29574 \times 10^{-5}$ & $-7.32805 \times 10^{-6} $\\
$u^{11}$ & $-3.69348 \times 10^{-9}$ & $-2.24398 \times 10^{-7}$ & $3.23179 \times 10^{-7}$ & $-1.06627 \times 10^{-6}$ & $5.58160 \times 10^{-7} $\\
$u^{12}$ & $1.63951 \times 10^{-10}$ & $9.71929 \times 10^{-9}$ & $-1.55738 \times 10^{-8}$ & $5.04247 \times 10^{-8}$ & $-2.47471 \times 10^{-8} $\\
$u^{13}$ & $-3.21841\times 10^{-12}$ & $-1.87011 \times 10^{-10}$ & $3.25850 \times 10^{-10}$ & $-1.04370 \times 10^{-9}$ & $4.84496 \times 10^{-10} $\\
\hline
$e^{-u}\times$ & & $S_{p}$ (1b+2b min) & $S_{p}$ (1b+2b max)& 
$S_{n}$ (1b+2b min) & $S_{n}$ (1b+2b max) \\
\hline
$1$&&$ 1.95620 \times 10^{-4}$ & $4.65803 \times 10^{-4}$ & $0.00601911$ & $0.00724841 $\\
$u$&&$ -3.13182 \times 10^{-4}$ & $-0.00101153$ & $-0.0166563$ & $-0.0200322 $\\
$u^2$&&$ 0.00203908$ & $0.00478674$ & $0.112020$ & $0.129805 $\\
$u^3$&&$ -0.00294523$ & $-0.00367367$ & $-0.152800$ & $-0.168826 $\\
$u^4$&&$ 0.00372071$ & $7.15246 \times 10^{-5}$ & $0.0969703$ & $0.0946747 $\\
$u^5$&&$ -0.00397760$ & $0.00103907$ & $-0.0375392$ & $-0.0254257 $\\
$u^6$&&$ 0.00277014$ & $-6.23246 \times 10^{-4}$ & $0.0107886$ & $0.00136908 $\\
$u^7$&&$ -0.00122679$ & $2.19242 \times 10^{-4}$ & $-0.00265967$ & $0.00140641 $\\
$u^8$&&$ 3.56712 \times 10^{-4}$ & $-5.63446 \times 10^{-5}$ & $5.62524 \times 10^{-4}$ & $-5.71526 \times 10^{-4} $\\
$u^9$&&$ -6.94095 \times 10^{-5}$ & $1.05667 \times 10^{-5}$ & $-9.21422 \times 10^{-5}$ & $1.20040 \times 10^{-4} $\\
$u^{10}$&&$ 8.97920 \times 10^{-6}$ & $-1.35391 \times 10^{-6}$ & $1.06409 \times 10^{-5}$ & $-1.58265 \times 10^{-5} $\\
$u^{11}$&&$ -7.41548 \times 10^{-7}$ & $1.10132 \times 10^{-7}$ & $-8.02220 \times 10^{-7}$ & $1.30826 \times 10^{-6} $\\
$u^{12}$&&$ 3.53473 \times 10^{-8}$ & $-5.11319 \times 10^{-9}$ & $3.53107 \times 10^{-8}$ & $-6.19338 \times 10^{-8} $\\
$u^{13}$&&$ -7.38713\times 10^{-10}$ & $1.03777\times 10^{-10}$ & $-6.87368\times 10^{-10}$ & $1.28102 \times 10^{-9} $\\
\hline
\end{tabular*}
\end{center}
\end{table}

\newpage

\onecolumngrid

\begin{table}[t]
\caption{Fits to the isoscalar/isovector structure factors $S_{00}$, 
$S_{11}$ and $S_{01}$ as well as ``proton-only" and ``neutron-only" 
structure factors $S_{p}$ and $S_{n}$ for spin-dependent WIMP
inelastic scattering off $^{131}$Xe nuclei, including 1b and 2b
currents as in Fig. 1. The upper and lower limits from the theoretical
error band were used for the fit.  The fitting function of the
dimensionless variable $u = p^2 b^2/2$ is $S_{ij}(u) = e^{-u}
\sum_{n=0}^{13} c_{ij,n} u^n$.  The rows give the coefficients
$c_{ij,n}$ of the $u^n$ terms in the polynomial.}
\begin{center}
\begin{tabular*}{0.795\textwidth}{c||c|c|c|c|c}
\hline
\multicolumn{6}{c}{$^{131}$Xe inelastic } \\
\multicolumn{6}{c}{$u=p^2b^2/2 \,, \: b=2.2905 \, {\rm fm}$} \\
\hline
$e^{-u}\times$ & $S_{00}$ & $S_{11}$ (1b+2b min) & $S_{11}$ (1b+2b max) & 
$S_{01}$ (1b+2b min) & $S_{01}$ (1b+2b max) \\
\hline
$1$ & $2.35664 \times 10^{-4}$ & $1.10853 \times 10^{-4}$ & $1.79755 \times 10^{-4}$ & $-4.15211 \times 10^{-4}$ & $-3.23966 \times 10^{-4} $\\
$u$ & $-6.08853 \times 10^{-4}$ & $-2.45643 \times 10^{-4}$ & $-4.33051 \times 10^{-4}$ & $0.00129716$ & $8.20866 \times 10^{-4} $\\
$u^2$ & $0.0755799$ & $0.0334167$ & $0.0530640$ & $-0.129644$ & $-0.101006 $\\
$u^3$ & $-0.101995$ & $-0.0558344$ & $-0.0902236$ & $0.205096$ & $0.153748 $\\
$u^4$ & $0.0557595$ & $0.0418290$ & $0.0707274$ & $-0.149803$ & $-0.102141 $\\
$u^5$ & $-0.0156606$ & $-0.0189589$ & $-0.0348228$ & $0.0707352$ & $0.0401662 $\\
$u^6$ & $0.00234686$ & $0.00598298$ & $0.0123283$ & $-0.0254224$ & $-0.0110710 $\\
$u^7$ & $-1.64837 \times 10^{-4}$ & $-0.00142758$ & $-0.00331303$ & $0.00729884$ & $0.00243568 $\\
$u^8$ & $-3.56864 \times 10^{-7}$ & $2.65780 \times 10^{-4}$ & $6.73999 \times 10^{-4}$ & $-0.00161524$ & $-4.49005 \times 10^{-4} $\\
$u^9$ & $1.19267 \times 10^{-6}$ & $-3.79453 \times 10^{-5}$ & $-1.00464 \times 10^{-4}$ & $2.60489 \times 10^{-4}$ & $6.57637 \times 10^{-5} $\\
$u^{10}$ & $-1.56233 \times 10^{-7}$ & $3.95240 \times 10^{-6}$ & $1.04819 \times 10^{-5}$ & $-2.91069 \times 10^{-5}$ & $-6.99980 \times 10^{-6} $\\
$u^{11}$ & $1.31250 \times 10^{-8}$ & $-2.79009 \times 10^{-7}$ & $-7.18478 \times 10^{-7}$ & $2.12299 \times 10^{-6}$ & $4.95896 \times 10^{-7} $\\
$u^{12}$ & $-6.37159 \times 10^{-10}$ & $1.18297 \times 10^{-8}$ & $2.89173 \times 10^{-8}$ & $-9.07671 \times 10^{-8}$ & $-2.07588 \times 10^{-8} $\\
$u^{13}$ & $1.35650\times 10^{-11}$ & $-2.26329 \times 10^{-10}$ & $-5.16254 \times 10^{-10}$ & $1.72399 \times 10^{-9}$ & $3.87968 \times 10^{-10} $\\
\hline
$e^{-u}\times$ & & $S_{p}$ (1b+2b min) & $S_{p}$ (1b+2b max)& 
$S_{n}$ (1b+2b min) & $S_{n}$ (1b+2b max) \\
\hline
$1$&&$ 5.65964 \times 10^{-6}$ & $2.28585 \times 10^{-5}$ & $6.70789 \times 10^{-4}$ & $8.32428 \times 10^{-4} $\\
$u$&&$ -8.66267 \times 10^{-5}$ & $-5.09691 \times 10^{-5}$ & $-0.00169273$ & $-0.00243141 $\\
$u^2$&&$ 0.00202856$ & $0.00813986$ & $0.210152$ & $0.258952 $\\
$u^3$&&$ 0.00339658$ & $-0.00456893$ & $-0.312067$ & $-0.398955 $\\
$u^4$&&$ -0.00915716$ & $-0.00374602$ & $0.200540$ & $0.278124 $\\
$u^5$&&$ 0.00811758$ & $0.00477228$ & $-0.0755643$ & $-0.122295 $\\
$u^6$&&$ -0.00416169$ & $-0.00227425$ & $0.0198702$ & $0.0404361 $\\
$u^7$&&$ 0.00143326$ & $6.58701 \times 10^{-4}$ & $-0.00421369$ & $-0.0108203 $\\
$u^8$&&$ -3.48016 \times 10^{-4}$ & $-1.34722 \times 10^{-4}$ & $7.63565 \times 10^{-4}$ & $0.00228304 $\\
$u^9$&&$ 5.95083 \times 10^{-5}$ & $2.03323 \times 10^{-5}$ & $-1.11246 \times 10^{-4}$ & $-3.56484 \times 10^{-4} $\\
$u^{10}$&&$ -6.95756 \times 10^{-6}$ & $-2.18809 \times 10^{-6}$ & $1.18177 \times 10^{-5}$ & $3.88612 \times 10^{-5} $\\
$u^{11}$&&$ 5.26485 \times 10^{-7}$ & $1.55368 \times 10^{-7}$ & $-8.36904 \times 10^{-7}$ & $-2.77670 \times 10^{-6} $\\
$u^{12}$&&$ -2.31918 \times 10^{-8}$ & $-6.46377 \times 10^{-9}$ & $3.50755 \times 10^{-8}$ & $1.16640 \times 10^{-7} $\\
$u^{13}$&&$ 4.51904\times 10^{-10}$ & $1.20019\times 10^{-10}$ & $-6.56349\times 10^{-10}$ & $-2.18138 \times 10^{-9} $\\
\hline
\end{tabular*}
\end{center}
\end{table}

\bibliographystyle{apsrev}
\bibliography{inelasticxenon}

\end{document}